\documentclass{article}

\usepackage{arxiv}
\usepackage[utf8]{inputenc}            
\usepackage[T1]{fontenc}               
\usepackage{amsmath}                   
\usepackage{hyperref}                  
\usepackage{url}                       
\usepackage{booktabs}                  
\usepackage{amsfonts}                  
\usepackage{nicefrac}                  
\usepackage{microtype}                 
\usepackage{cleveref}                  
\usepackage{authblk}                   
\usepackage{graphicx}
\usepackage{caption}                   
\usepackage{subcaption}                
\usepackage{float}                     
\usepackage[nolist]{acronym}           
\usepackage{multirow}                  
\usepackage{makecell}                  
\usepackage{siunitx}                   
\usepackage[dvipsnames, table]{xcolor} 
\usepackage[numbers, sort&compress]{natbib} 
\usepackage{enumitem}                  
\usepackage{doi}
\usepackage{subfiles}                       

\newcommand{\integer}[1]{\num[round-mode=places, round-precision=0, group-minimum-digits=4]{#1}}
\newcommand{\round}[1]{\num[round-mode=places, round-precision=3, group-minimum-digits=4]{#1}}

\newcommand{\beginsupplement}{
    \appendix
    \newpage
    \setcounter{table}{0}
    \renewcommand{\thetable}{S\arabic{table}}
    \renewcommand{\theHtable}{S\arabic{table}} 
    \setcounter{figure}{0}
    \renewcommand{\thefigure}{S\arabic{figure}}
    \renewcommand{\theHfigure}{S\arabic{figure}}
    \setcounter{section}{0}
    \renewcommand{\thesection}{S\arabic{section}}
    \renewcommand{\theHsection}{S\arabic{section}}
    \setcounter{equation}{0}
    \renewcommand{\theequation}{S\arabic{equation}}
    \renewcommand{\theHequation}{S\arabic{equation}}
    \setcounter{footnote}{0}
    \acresetall 
}

\title{Experimental Online Quantum Dots Charge Autotuning using Neural Networks}

\newbox{\orcid}\sbox{\orcid}{\includegraphics[scale=0.06]{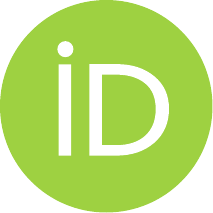}} 

\author[1,2,3]{%
    \href{https://orcid.org/0000-0003-4517-5042}{\usebox{\orcid}\hspace*{1mm}Victor~Yon
    \thanks{\texttt{victor.yon@usherbrooke.ca}}}%
}
\author[1,2,3]{ \href{https://orcid.org/0009-0005-0384-1109}{\usebox{\orcid}\hspace*{1mm}Bastien~Galaup} }
\author[2,3,4]{ \href{https://orcid.org/0009-0003-5789-3807}{\usebox{\orcid}\hspace*{1mm}Claude~Rohrbacher} }
\author[2,3,4]{ Joffrey~Rivard }
\author[2,3,4]{ Alexis~Morel }
\author[2,3,4]{ \href{https://orcid.org/0009-0003-8185-4998}{\usebox{\orcid}\hspace*{1mm}Dominic~Leclerc} }
\author[5]{ \href{https://orcid.org/0000-0002-5244-3474}{\usebox{\orcid}\hspace*{1mm}Clément~Godfrin} }
\author[5]{ \href{https://orcid.org/0000-0002-2145-7590}{\usebox{\orcid}\hspace*{1mm}Ruoyu~Li} }
\author[5]{ Stefan~Kubicek }
\author[5]{ \href{https://orcid.org/0000-0002-1314-9715}{\usebox{\orcid}\hspace*{1mm}Kristiaan~De~Greve} }
\author[2,3,4]{ Eva~Dupont-Ferrier }
\author[1,2,3]{ \href{https://orcid.org/0000-0003-0311-8840}{\usebox{\orcid}\hspace*{1mm}Yann~Beilliard} }
\author[6,7]{ \href{https://orcid.org/0000-0002-5505-8176}{\usebox{\orcid}\hspace*{1mm}Roger~G.~Melko} }
\author[1,2,3]{ \href{https://orcid.org/0000-0003-2156-967X}{\usebox{\orcid}\hspace*{1mm}Dominique~Drouin} }

\affil[1]{Institut Interdisciplinaire d'Innovation Technologique (3IT), Université de Sherbrooke, Sherbrooke, QC,
    Canada, J1K~0A5}
\affil[2]{Laboratoire Nanotechnologies Nanosystèmes (LN2) — CNRS 3463, Université de Sherbrooke, Sherbrooke, QC, Canada,
    J1K~0A5}
\affil[3]{Institut quantique (IQ), Université de Sherbrooke, Sherbrooke, QC, Canada, J1K~2R1}
\affil[4]{Département de physique, Université de Sherbrooke, Sherbrooke, QC, Canada, J1K~2R1}
\affil[5]{IMEC, Kapeldreef 75, 3001 Leuven, Belgium}
\affil[6]{Department of Physics and Astronomy, University of Waterloo, Waterloo, ON, Canada, N2L~3G1}
\affil[7]{Perimeter Institute for Theoretical Physics, Waterloo, ON, Canada, N2L~2Y5}

\renewcommand{\shorttitle}{Experimental Online QDs Charge Autotuning using NNs}

\hypersetup{
    pdftitle={Experimental online quantum dots charge autotuning using neural networks},
    pdfsubject={cs.AI, cs.LG, quant-ph},
    pdfauthor={Victor Yon, Bastien Galaup, Claude Rohrbacher, Joffrey Rivard, Alexis Morel, Dominic Leclerc,
    Clement Godfrin, Ruoyu Li, Stefan Kubicek, Kristiaan De Greve, Eva Dupont-Ferrier, Yann Beilliard,
    Roger G. Melko, Dominique Drouin},
    pdfkeywords={quantum dot, spin qubit, machine learning, convolutional neural network, scalable quantum computing,
    autonomous calibration, charge autotuning},
}

\begin{document}
    \maketitle

    \begin{abstract}

Spin-based semiconductor qubits hold promise for scalable quantum computing, yet they require reliable autonomous calibration procedures.
This study presents an experimental demonstration of online single-dot charge autotuning using a convolutional neural network integrated into a closed-loop calibration system.
The autotuning algorithm explores the gates' voltage space to localize charge transition lines, thereby isolating the one-electron regime without human intervention.
This exploration leverages the model's uncertainty estimation to find the appropriate gate configuration with minimal measurements while reducing the risk of failures.
In \num{20} experimental runs, our method achieved a success rate of \qty{95}{\percent} in locating the target electron regime, highlighting the robustness of this approach against noise and distribution shifts from the offline training set.
Each tuning run lasted an average of \num{2} hours and \num{9} minutes, primarily due to the limited speed of the current measurement.
This work validates the feasibility of machine learning-driven real-time charge autotuning for quantum dot devices, advancing the development toward the control of large qubit arrays.

    \end{abstract}

    \keywords{
        quantum dot \and
        spin qubit \and
        machine learning \and
        convolutional neural network \and
        scalable quantum computing \and
        autonomous calibration \and
        charge autotuning
    }


    \begin{acronym}
        \acro{CMOS}{complementary metal-oxide-semiconductor}
        \acro{CNN}{convolutional neural network}
        \acro{DNN}{deep neural network}
        \acro{FF}{feed-forward neural network}
        \acroindefinite{FF}{an}{a}
        \acro{ML}{machine learning}
        \acroindefinite{ML}{an}{a}
        \acro{NN}{neural network}
        \acroindefinite{NN}{an}{a}
        \acro{QD}{quantum dot}
        \acro{SEM}{Scanning electron microscope}
        \acro{SET}{single-electron transistor}
        \acro{Si-OG-QD}{silicon overlapping gates quantum dot}
        \acro{FPGA}{field-programmable gate array}
        \acroindefinite{FPGA}{an}{a}
    \end{acronym}


    \section{Introduction}\label{sec:introduction}

Semiconductor spin qubits~\cite{Loss_1998, Veldhorst_2015, Watson_2018, Burkard_2023, Zwanenburg_2013} can encode quantum information using the spin-\nicefrac{1}{2} of a charge carrier, which can be manipulated using external magnetic fields to perform quantum computing based on the principles of superposition and entanglement.
This technology stands out due to its high gate fidelity~\cite{Takeda_2016, Yoneda_2017, Mills_2022, Noiri_2022, Xue_2022, Weinstein_2023, Gilbert_2023}, long coherence times~\cite{Tyryshkin_2011, Veldhorst_2014}, thermal robustness~\cite{Petit_2022, Yang_2020, Huang_2024}, and compatibility with existing \ac{CMOS} technologies~\cite{Maurand_2016, Stuyck_2021, Zwerver_2022, Rochette_2019, Rohrbacher_2023}.
These characteristics make spin qubits excellent candidates for building scalable quantum computers using already existing industrial fabrication methods~\cite{Gonzalez_2021, Rohrbacher_2023, Neyens_2024, Rohrbacher_2024}.
However, significant engineering challenges remain, such as improving device fabrication quality~\cite{Dodson_2020, Tahan_2021, Michniewicz_2024, Saraiva_2021} and developing autonomous calibration procedures for a large number of \acp{QD}~\cite{Borsoi_2023, Mouny_2024}.

A spin qubit is formed by trapping a specific number of charge carriers within an isolated island (\ac{QD}) using, for example, the electrostatic confinement gates of a nanoscale device (as shown in Figure~\ref{fig:setup}a).
Calibrating the \ac{QD} device to achieve a specific physical state is a complex task that is generally approached in a series of sequential steps~\cite{Zwolak_2023}.
Initially, the device is cooled, local charge sensing systems are activated, and the voltages of the confinement gates are adjusted to operate within appropriate ranges for data collection~\cite{Zubchenko_2024, Kovach_2024}.
Following this, the gate voltage ranges corresponding to a known global structure (such as single- or double-\ac{QD} configurations) are calibrated.
This step is usually referred to as coarse tuning~\cite{Ziegler_2023, Liu_2022, Zwolak_2020, Kalantre_2019, Darulova_2021, Moon_2020, Ziegler_2022, Severin_2024}.
Next, optional virtual gates could be established to compensate for capacitive crosstalk, ensuring precise control of the individual \acp{QD} without unwanted interference~\cite{Ziegler_2023, Liu_2022, Perron_2015, Hensgens_2018, Hsiao_2020, vanDiepen_2018, Rao_2024}.
Subsequently, the device gate voltages are precisely tuned to achieve a specific charge carrier count in each \ac{QD}, which in our experiment refers to the number of electrons in a single \ac{QD}.
This step is usually referred to as charge state tuning~\cite{Ziegler_2023, Czischek_2021, Lapointe_2020, Durrer_2020, Yon_2024, Muto_2024, Baart_2016, Meyer_2023}.
Finally, the system's physical parameters are refined in preparation for quantum computations~\cite{Liu_2022, Teske_2019, Botzem_2018, Daraeizadeh_2020}.

Performing efficient and accurate calibration every time the system is cooled down is critical to deploying large-scale \ac{QD}-based systems.
However, automatizing this process is challenging due to the sensitivity of the operational parameters, where each variable can affect others nonlinearly, exponentially increasing the complexity as the number of tuned devices grows.
Variability in device fabrication adds another layer of difficulty, as each \ac{QD} can behave differently, requiring customized tuning approaches.
These devices and the measurement method are also susceptible to environmental conditions (e.g., thermal noise and electromagnetic interference), adding stochasticity to the measurements.

Recent progress in producing larger arrays of \acp{QD}~\cite{Borsoi_2023, Neyens_2024} has accelerated the need for robust control procedures.
This challenge has been partially addressed by leveraging \ac{ML} models~\cite{Kalantre_2019, Moon_2020}, which can be used to tune quantum devices by automatically navigating through the large and noisy parameter space.
However, \ac{ML} methods, especially \acp{NN}, are sensitive to the distribution shift~\cite{Gawlikowski_2021} between the training and testing data and are well known for unexpected failures~\cite{Goodfellow_2014}.
Therefore, it is necessary to validate any \ac{NN}-based autotuning method in a real-world experimental environment~\cite{Zubchenko_2024}.
However, due to the scarcity and expense of the hardware required to run such experiments, most autotuning demonstrations are performed offline (i.e., using static pre-recorded data).
A few studies have demonstrated online experimental procedures for the first steps of the autotuning process: bootstrapping~\cite{Zubchenko_2024}, coarse tuning~\cite{Severin_2024, Zwolak_2020}, and creating virtual gates~\cite{vanDiepen_2018, Hsiao_2020}.
\citet{Baart_2016} showed an early demonstration of charge state autotuning using Gabor filters~\cite{Mehrotra_1992} and classical optimization algorithms, resulting in a single successful run after \num{3} hours.
Only \citet{Schuff_2024} ran and benchmarked online experiments that covered the full tuning procedure.
Their method---based on a combination of tree search, Bayesian optimization, segmentation algorithms, and \acp{NN}---allowed them to reach \qty{77}{\percent} success over \num{13} runs for complete \ac{QD} tuning.
However, each run took an average of \num{38} hours to complete due to the time required to measure large stability diagrams and the repeated calibration necessary after a stage failure.

The autotuning algorithm used during the experiment has been developed and tested offline in a previous study~\cite{Yon_2024}.
It relies on a \ac{CNN}~\cite{Krizhevsky_2017, OShea_2015, Patil_2020} trained to detect charge transition lines in a small section of the stability diagram.
The training is performed in a supervised manner on a dataset~\cite{qdsd} composed of static measurements made on similar silicon \ac{QD} devices.
Then, a closed-loop calibration procedure allows us to find the one-electron regime of a single \ac{QD} by following an autonomous exploration strategy that leverages the \ac{CNN} inference and uncertainty score.
Exploiting the model's uncertainty improves the robustness of the autotuning by reducing the risk of critical failures caused by potential misclassifications~\cite{Yon_2024} compared to classical \ac{ML} methods~\cite{Czischek_2021, Zwolak_2023}.

In this experiment, we successfully transfer the above \ac{ML}-based charge autotuning method---developed using offline measurements---to real-time charge tuning on an experimental setup.
The results confirm that the device-to-device variability and the resolution shift between the online and offline data do not negatively affect the line detection performance.
We were also able to measure valuable information regarding the autotuning duration and identify the current measurement as the time bottleneck.

    \section{Methodology}\label{sec:methodology}

\begin{figure}
    \centering
    \includegraphics[width=.95\textwidth]{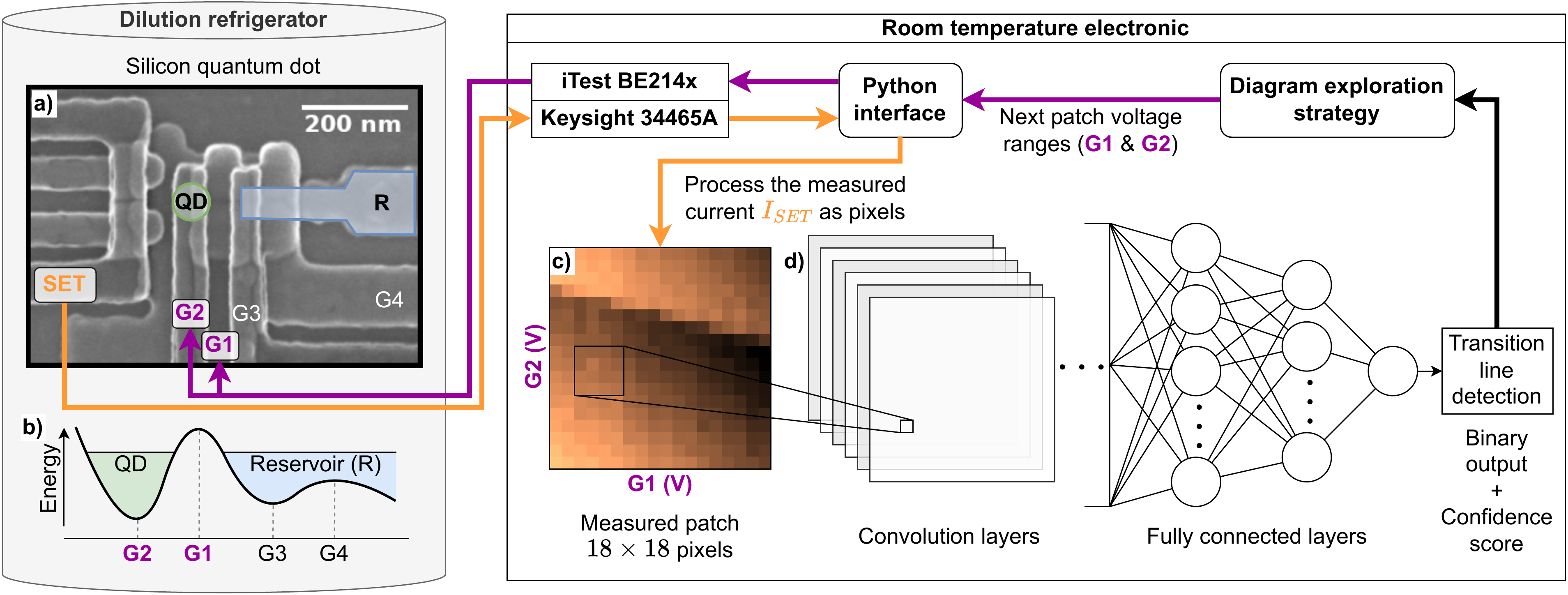}
    \caption[Schematic representation of the experimental setup]{
        Schematic representation of the experimental setup.
        A  silicon \acf*{QD} device is cooled in a dilution refrigerator with a base temperature of \qty{20}{\milli\kelvin}.
        The voltages at gates \num{3} (G3) and \num{4} (G4) are fixed, while a \emph{iTest BE214x} controls the voltages at gates \num{1} (G1) and \num{2} (G2) through a Python interface.
        The scan coordinates are defined by an autonomous closed-loop procedure using a trained \acf*{CNN} and an exploration strategy algorithm.
        \textbf{a)}~\acf*{SEM} image of a silicon \ac{QD} device with overlapping gates, similar to the one used during this experiment.
        The electrons flow from the reservoir (R) to the \ac{QD}.
        See~\cite{Elsayed_2024} for fabrication details.
        \textbf{b)}~Energy diagram representing the formation of a single \ac{QD} in this device.
        \textbf{c)}~A subsection of the voltage space (referred to as ``patch'') was scanned by sweeping the voltages at gates G1 and G2 and measuring the current using the \acf*{SET} connected to a \emph{Keysight 34465A} multimeter.
        \textbf{d)}~Trained \ac{CNN} that processed the measured patch through two convolutional layers and two fully connected layers to infer the presence of a transition line as a binary output.
    }
    \label{fig:setup}
\end{figure}

The charge tuning process is typically guided by a stability diagram (two-dimensional current--voltage scan presented in Figure~\ref{fig:results}a), which is generated based on indirect \ac{QD} measurements from \iac{SET} for charge detection while sweeping the gate voltages.
The transition lines (highlighted in green in Figure~\ref{fig:results}c) inform us of an electron movement between the reservoir and the \ac{QD} (represented in Figure~\ref{fig:setup}a,b).
Given the knowledge that the \ac{QD} is empty when no lines are visible at low gate voltages, it is possible to deduce the number of charges for a given position in the stability diagram (blue areas in Figure~\ref{fig:results}c).
Although experts can perform manual tuning in small-scale experiments~\cite{Meyer_2023}, this is too slow and labor-intensive for large-scale industrial applications utilizing multiple \acp{QD}.
Automating this task is necessary but challenging due to the noise induced by the environment and the measurement electronics, the device-to-device variability, and the variety of existing hardware implementations.

\begin{figure}
    \centering
    \hfill
    \begin{subfigure}[c]{0.30\textwidth}
        \centering
        \includegraphics[width=\textwidth]{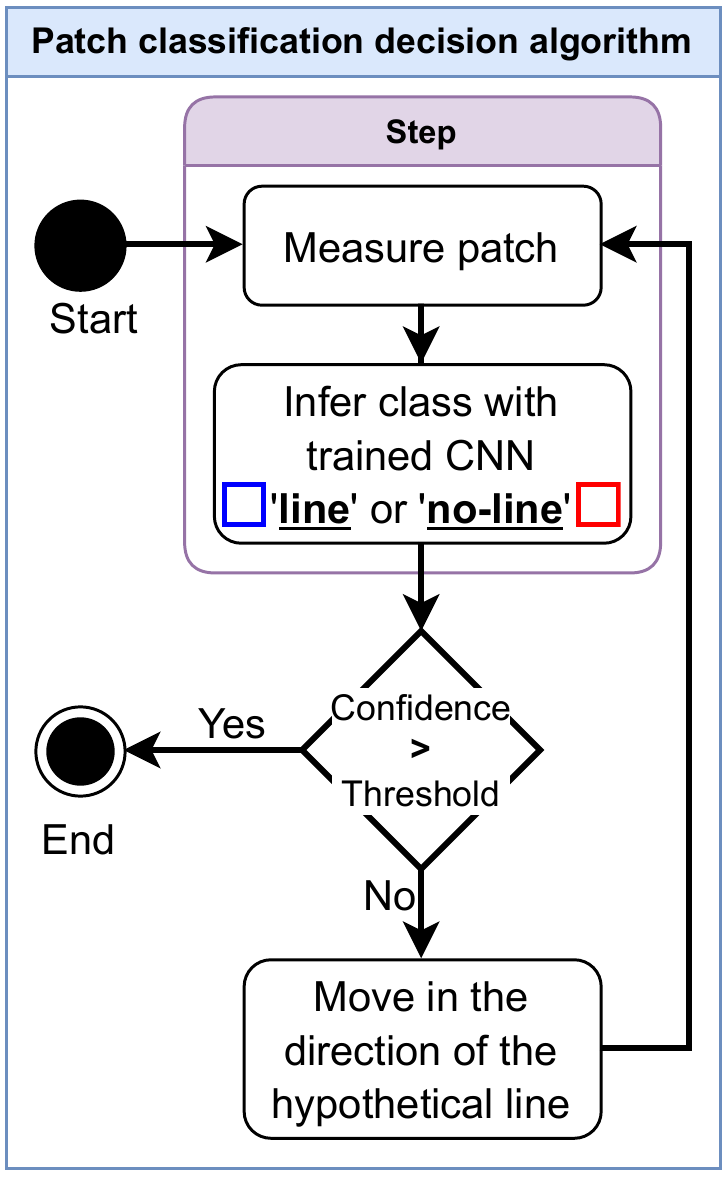}
        \caption{Patch classification logic represented as a flow diagram.}
        \label{fig:classification-flow}
    \end{subfigure}
    \hfill
    \begin{subfigure}[c]{0.65\textwidth}
        \centering
        \includegraphics[width=\textwidth]{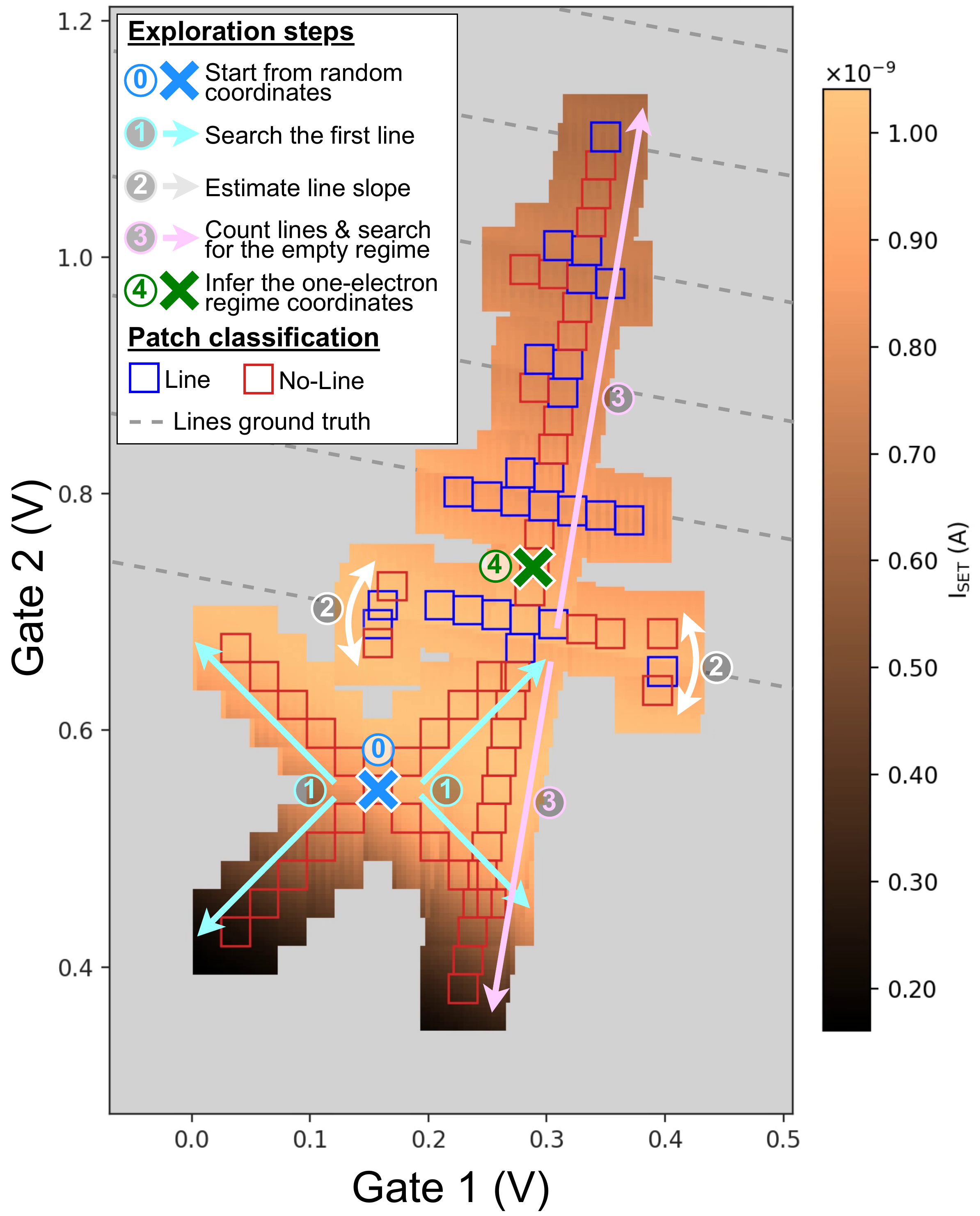}
        \caption{Example of a successful online experimental autotuning.}
        \label{fig:tuning-example}
    \end{subfigure}
    \hfill
    \caption[Example of successful online experimental autotuning]{
        Autonomous exploration strategy.
        One patch classification step of the algorithm is described in the subfigure~\textbf{(a)}.
        Multiple iterations of this step are represented by the blue and red squares in the subfigure~\textbf{(b)}.
        In which, the arrows represent the direction of the exploration, and the gray area represents the unmeasured voltage space for this run.
        The cyan arrows \emph{(1)}~represent the exploration in four directions to search for the first transition line.
        The white arrows \emph{(2)}~represent the slope estimation step, performed by scanning two sections of the first detected line.
        Finally, the pink arrows \emph{(3)}~perpendicular to the estimated line slope represent the empty regime search \emph{(down)} and the line count \emph{(up)} procedure.
        A complete scan of this diagram is given in Figure~\ref{fig:results}a.
        See Supplementary Section~\ref{sec:suppl-exploration-strategy} for more detail regarding this autonomous exploration method.
    }
    \label{fig:tuning-method}
\end{figure}

The autotuning is approached as an exploration problem, where we start from a random unknown position in the voltage space, gather information about the surroundings by performing local measurements, and search for the targeted charge regime.
The transition lines guide the exploration by providing valuable information regarding the number of electrons in the \ac{QD}.
One exploration step consists of scanning a subsection of the voltage space (referred to as ``patch'' and represented in Figure~\ref{fig:setup}c), sending it to the input of a \ac{CNN}-based line detection model and deciding the next area to explore based on an exploration strategy (see Figure~\ref{fig:tuning-method}).
The patch size is fixed to \numproduct{18 x 18} data points as a tradeoff between the measurement time (a smaller area is faster to scan) and the line detection accuracy (a larger area simplifies the line detection).
Due to the noisy nature of the measurements and the small size of the patch, classical signal processing and pattern detection methods~\cite{Sun_2022, Mukhopadhyay_2015, Ding_2001, Bishop_2006} are not robust enough to reliably detect transition lines~\cite{Czischek_2021, Baart_2016, Lapointe_2020, Ziegler_2023}.
We opted for a supervised classification approach using a \ac{DNN}~\cite{LeCun_2015}, which is known to be the best-performing method for pattern detection in noisy images~\cite{Voulodimos_2018, Cai_2020, Krizhevsky_2017, OShea_2015, Patil_2020}.

The line detection and exploration strategy are based on the autotuning method described in detail by \citet{Yon_2024} (see Supplementary Section~\ref{sec:suppl-exploration-strategy}).
Nine stability diagrams, measured during previous experiments on similar \ac{QD} devices and manually annotated, are used to generate a training set of \num{33429} patches.
Each of these patches is categorized as ``\emph{line}'' or ``\emph{no-line}'' depending on whether an annotation of a transition line intersects its center.
More information on the diagram annotation and patch labeling is available in Supplementary Section~\ref{sec:suppl-datasets}.
A \ac{CNN} (represented in Figure~\ref{fig:setup}d) is then optimized using gradient backpropagation~\cite{Rojas_1996} to classify patches in the training set (refer to Supplementary Section~\ref{sec:suppl-training} for details on the training methodology).

The trained \ac{CNN} is transferred to a computer connected to a \emph{iTest BE214x} that is capable of controlling the voltage applied to each gate of the \ac{QD} device and measuring the \ac{SET} current using a \emph{Keysight 34465A} multimeter, as illustrated in Figure~\ref{fig:setup}.
The autotuning algorithm, implemented in Python\footnote{Python source code: \href{https://github.com/3it-inpaqt/dot-calibration-v2}{github.com/3it-inpaqt/dot-calibration-v2}}, plays a central role at each step of the exploration by: \emph{(i)}~determining the next patch to measure based on the exploration strategy, \emph{(ii)}~transferring the voltage sweeping instructions to the multimeter, \emph{(iii)}~processing the measured current as a normalized image, and \emph{(iv)}~detecting a transition line by feeding the measured patch to the \ac{CNN} input.
To reduce the risk of tuning failures induced by potential patch misclassifications, each inference is associated with a confidence score that estimates the model's uncertainty for a given input.
This score is calculated using a simple distance-based heuristic~\cite{Zaragoza_1998, Mandelbaum_2017} (more information in Supplementary Section~\ref{sec:suppl-confidence-score}).
When a patch is classified with a confidence score under the threshold, a verification procedure is automatically triggered to validate or refute the presence of a line in the surrounding area.
This uncertainty-based exploration has been demonstrated to significantly improve the tuning success rate in offline experiments~\cite{Yon_2024}.
It is therefore expected to improve the robustness of autotuning in the case of unexpected perturbations related to the experimental context.

A silicon \ac{QD} device with overlapping gates~\cite{Elsayed_2024} was installed and cooled down in a dilution refrigerator with a base temperature of \qty{20}{\milli\kelvin}.
Once the device was cooled, we manually configured it in a single-\ac{QD} state by keeping the voltages of the confinement gates and unused gates at \qty{0}{\V}, while applying \qty{3}{\V} to the gates we wanted to accumulate.
These values were determined through standard I--V measurements.
We then ran \num{20} consecutive and independent charge autotuning procedures.

Each run started at a random location (blue crosses in Figure~\ref{fig:results}b) in a voltage range of [\qty{-0.25}{\V},~\qty{0.5}{\V}] for gate \num{1} (G1) (working range for the barrier gate between the \ac{QD} and the reservoir, based on pinch-off measurement) and [\qty{0}{\V},~\qty{2}{\V}] for gate \num{2} (G2) (working range for a single \ac{QD} based on I--V charactrisations).
Defining bounds on the gates' voltages was necessary to maintain the hardware's integrity and avoid wasting time on regions incompatible with a single-\ac{QD} configuration.
However, since the boundary values are generally consistent across all samples of the same type, characterization is required only on the first device.

The voltage space was then explored using iterative patch measurements within the starting voltage ranges, extended by a \qty{0.1}{\V} margin in all directions to ensure the exploration did not begin near a boundary.
We did not set virtual gates, as the tuning algorithm was designed to adapt to the capacitive crosstalk.
At the end of the experimental runs, we performed a complete scan of the explored area.
The resulting stability diagram (shown in Figure~\ref{fig:results}a) was annotated by experts to identify the transition lines and the charge areas (shown in Figure~\ref{fig:results}c).
By placing the final coordinates of the \num{20} runs in this figure, we were able to count the number of times we reached the targeted one-electron regime.

    \section{Results}\label{sec:results}

\begin{figure}
    \centering
    \includegraphics[width=.95\textwidth]{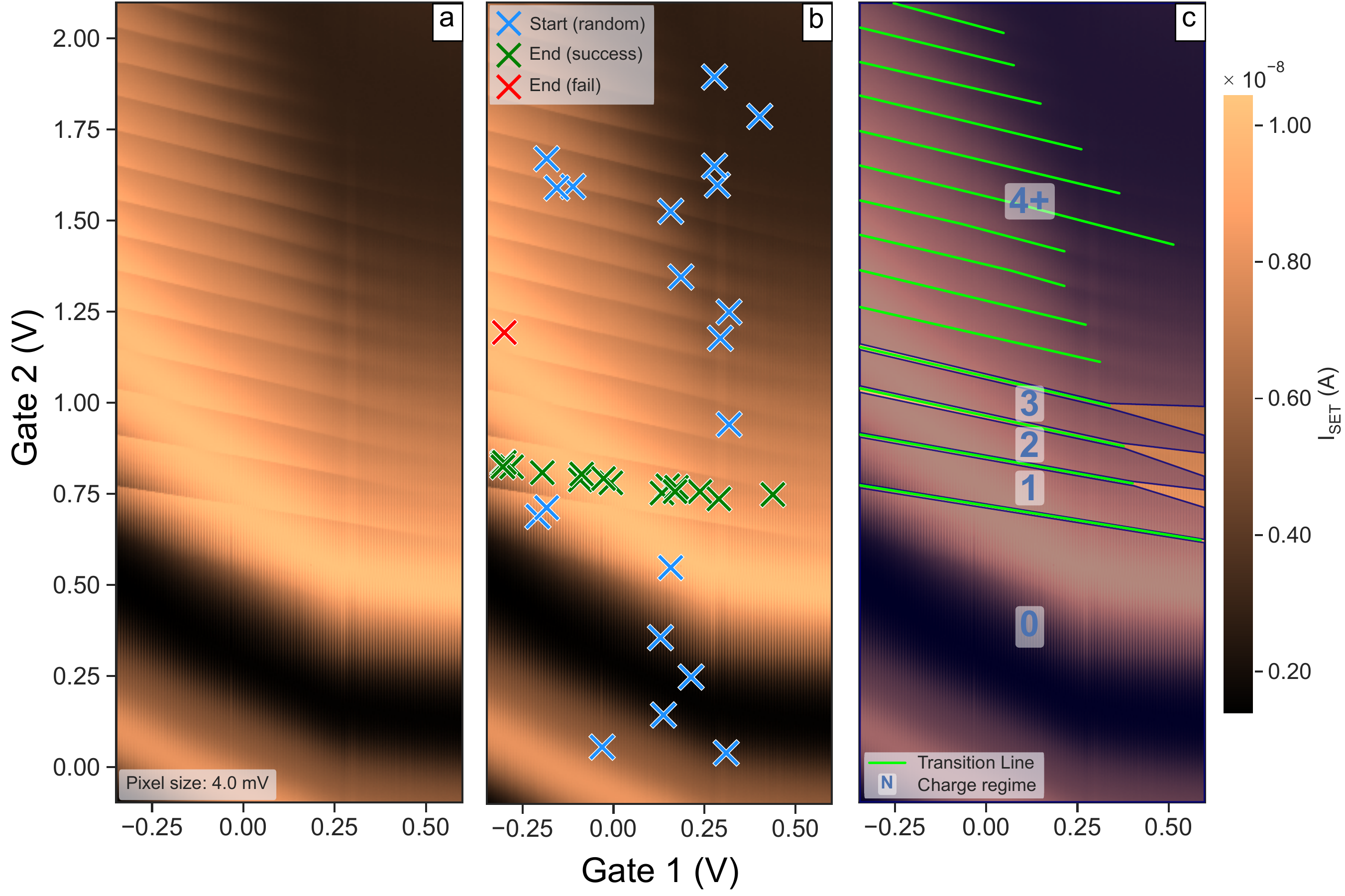}
    \caption[Complete scan of the explored stability diagram, processed after the experiment]{
        Complete scan of the explored stability diagram, processed after the experiment.
        \textbf{a)}~Representation of a stability diagram as an image, where pixel values encode the current measured using the \acf*{SET} as a function of the gate voltages applied to the \acf*{QD} (see Figure~\ref{fig:setup}).
        \textbf{b)}~Start and end coordinates of the \num{20} online experimental autotuning runs.
        The blue crosses represent the random starting point of the runs.
        The green crosses represent the final coordinates of the successful autotuning runs (when the coordinates are inside the area annotated as a one-electron regime).
        The red cross represents the end coordinates of the run that did not reach the target regime.
        \textbf{c)}~Same diagram with manual annotations of transition lines in green and charge regime areas in blue.
        The region with four or more charges is annotated as ``\emph{4+}''.
        The voltage areas not covered by a charge annotation (due to fading lines) are considered ``\emph{unknown charge regimes}''.
    }
    \label{fig:results}
\end{figure}

Positioning the final voltage coordinates in a complete scan of the explored stability diagram (Figure~\ref{fig:results}a) allows us to evaluate which autotuning runs reach the targeted charge regime.
Among the \num{20} experimental tunings performed, \num{19} (\qty{95}{\percent}) successfully located the one-electron regime (green crosses in Figure~\ref{fig:results}b), while only \num{1} (\qty{5}{\percent}) autotuning run failed (red cross in Figure~\ref{fig:results}b).
An analysis of this failure revealed that it was not caused by patch misclassification but by a problem in the exploration logic related to the voltage boundaries.
This issue has been fixed in the last version of the Python implementation.
A visual animation of each run is available in a video\footnote{Online autotuning experiment video: \href{https://youtu.be/zGIQWEZex0s}{youtu.be/zGIQWEZex0s}}.

Each autotuning run took an average of \num{2} hours and \num{9} minutes (standard deviation: \qty{46}{\minute}) and \num{110} steps (standard deviation: \num{38}) to complete.
For comparison, scanning the full stability diagram presented in Figure~\ref{fig:results}a took approximately \num{7} hours.
Each step required \num{324} current measurements to obtain one patch (for \numproduct{18 x 18} pixels), taking an average of \num{67} seconds using a \emph{Keysight 34465A} multimeter, representing \qty{96}{\percent} of the process duration.
Thus, data transfer and processing (including the \ac{CNN} inference time) represent only a fraction of the tuning time.
The high variability between run durations is attributable to the variable distance from the random starting point to the one-electron regime.
The individual run statistics are available in Supplementary Table~\ref{tab:runs-stats}.

Prior offline benchmarks obtained by applying this autotuning method to similar devices led to a lower tuning success rate (\qty{78}{\percent})~\cite{Yon_2024} despite the comparable line detection accuracy of the \acp{CNN}.
The higher success rate obtained during this online experiment can be explained by several factors:
\emph{(i)}~Online tuning allows us to precisely select the patch resolution, while offline diagrams need to be interpolated to compensate for the non-homogeneous step size of the voltage sweeping between measurements (see comparison examples in Supplementary Section~\ref{sec:suppl-patch-examples}).
This data processing is detrimental to the measurement quality and could explain some line detection failures during offline tests.
\emph{(ii)}~The fixed voltage range covered by offline diagrams often artificially increases the tuning difficulty by allowing for the exploration of irrelevant areas (e.g., beyond the barrier threshold).
\emph{(iii)}~Offline autotuning experiments were performed on data measured from multiple devices, covering an extensive range of physical defects and measurement noise.
The unique device used in this online experiment appears to be of a good quality (e.g., low noise, no parasitic dots), providing optimal autotuning conditions.

    \section{Discussion}\label{sec:discussion}

This online experiment demonstrates the feasibility of autonomous real-time tuning based on \ac{ML}.
It validates that the expected distributional shift between the offline training data and the online measurements is not detrimental to the procedure's performance.
On the contrary, the higher success rate observed in this experiment, compared to previous offline benchmarks on similar devices~\cite{Yon_2024}, suggests that the absence of preprocessing improved the robustness of the autotuning.
We also confirmed that the line detection model and the exploration strategy were versatile enough to successfully tune a \ac{QD} device that was not in the training set.

The relatively slow tuning time (\num{>2} hours) does not satisfy the requirement for practical tuning of a large \ac{QD} array.
However, the \ac{SET} measurement time, identified as the main bottleneck, could be drastically improved by optimizing the sensing method and hardware.
For example, one could speed up the measurement time by a factor of \num{10} by implementing the measurement sequence on a dedicated processor~\cite{Stehlik_2015} (e.g., \iac{FPGA}).
This approach would cut the communication time between the instruments and the control computer, allowing for faster sweep rates.
The radio-frequency reflectometry technique~\cite{Liu_2021, Zubchenko_2024} could also be used to speed up the charge sensing by measuring the impedance of the \ac{SET} sensor at a fixed frequency, enabling single-shot readouts with only several microseconds of integration time~\cite{Reilly_2007}.
This time can even be reduced to \qty{400}{\nano\second} with a Josephson parametric amplifier that also provides a high signal-to-noise ratio~\cite{Stehlik_2015}.
However, the whole measurement pipeline should be optimized to take advantage of this hardware acceleration, including the data transfer, latency, and processing time.
On the other hand, it is possible to limit the effect of this measurement bottleneck with a software approach; for example, by optimizing the meta-parameters (Supplementary Tables~\ref{tab:model-achitecture} and~\ref{tab:meta-parameters}) in a way that reduces the number of measured points (e.g., smaller patches and a larger voltage distance between pixels) and reducing the number of steps (e.g., higher confidence threshold and smaller voltage boundaries).
Finally, improving the fabrication processes of the \ac{QD} chip is likely to reduce device-to-device variability, which will directly improve the autotuning robustness and reduce the number of steps by narrowing the voltage ranges to explore.

Further optimization can be achieved by moving the control electronics (represented in the right panel of Figure~\ref{fig:setup}) to the cryogenic environment inside the dilution refrigerator.
Bringing the control electronics closer to the tuned device, either by co-integrating them with the \ac{QD} chip~\cite{Rohrbacher_2023, Rohrbacher_2024} or with additional nearby electronics, can reduce the parasitic capacitance added to the current measurements and allow for the control of large-scale arrays of \acp{QD} by circumventing the wiring bottleneck~\cite{Reilly_2019} between the \ac{QD} devices and the control electronics.
However, this \emph{in situ} autotuning scheme requires the development of cryo-compatible and low-power custom electronics~\cite{Mouny_2023, Yon_2024, Dawant_2024, Xue_2021, Yon_2024b}.


\section*{Acknowledgments}\label{sec:acknowledgments}

V.Y. acknowledges Christian Lupien's valuable technical assistance in interfacing the experimental hardware with Python.

V.Y., B.G., Y.B., and D.D. acknowledge support from the National Science Engineering Research Council of Canada, Grant ALLRP 580722--22, and the Fonds de Recherche du Québec---Nature et Technologies, Grant 300253.

C.R., J.R., A.M., D.L., and E.D.F acknowledge support from the FRQNT établissement de la relève professorale, Grant 2020--NC--268397, and the CRSNG, Grant RGPIN--2020--0573.

R.G.M. acknowledges support from NSERC and the Perimeter Institute for Theoretical Physics.
Research at the Perimeter Institute is supported in part by the Government of Canada through the Department of Innovation, Science and Economic Development Canada and by the Province of Ontario through the Ministry of Economic Development, Job Creation and Trade.

\section*{Conflicts of interest}\label{sec:conflict}

The authors declare no conflicts of interest.

\section*{Author contributions}\label{sec:contributions}

All authors contributed to this article and approved of the submitted version.

Victor Yon: methodology, software implementation, experiments, results analysis and visualization, writing--original draft preparation.

Bastien Galaup: software implementation, experiments, manuscript review.

Claude Rohrbacher, Joffrey Rivard, Alexis Morel, and Dominic Leclerc: initialize and configure the experimental setup, methodology, manuscript review.

Clement Godfrin, Ruoyu Li, Stefan Kubicek, and Kristiaan De Greve: manufactured the devices, manuscript review.

Roger Melko: supervision, manuscript review.

Eva Dupont-Ferrier, Yann Beilliard, and Dominique Drouin: methodology, supervision, funding acquisition, manuscript review, editing.

\section*{Code and data availability}\label{sec:data}

The Python source code used to aggregate and build the dataset is publicly accessible on \emph{GitHub}: \href{https://github.com/3it-inpaqt/qdsd-dataset}{github.com/3it-inpaqt/qdsd-dataset}.

The Python source code used to run all the experiments presented in this article is publicly accessible on \emph{GitHub}: \href{https://github.com/3it-inpaqt/dot-calibration-v2}{github.com/3it-inpaqt/dot-calibration-v2}.

The raw and processed stability diagram measurements used to train the line detection model for this experiment are publicly available for download from \citet{qdsd}.
The subset used to train the models in this paper is referred to as \ac{Si-OG-QD}.

All trained models, experimental measurements, and results presented in this article are publicly available for download from \citet{run_outputs_online}.

    \newpage
    \bibliographystyle{unsrtnat}
    \bibliography{references}

    \beginsupplement

\begin{center}
    \textbf{\huge Supplementary Materials: Experimental Online Quantum Dots Charge Autotuning using Neural Networks}
\end{center}

\section{Datasets}\label{sec:suppl-datasets}

The dataset~\cite{qdsd} is built from nine offline stability diagrams measured during previous experiments on similar \ac{QD} devices.
Each of them was manually annotated by experts with transition lines and charge regime areas (example in Figure~\ref{fig:setup}c).
The stability diagrams were decomposed into \num{33429} patches of \numproduct{18 x 18} pixels, then split into three subsets:

\begin{itemize}
    \item \qty{80}{\percent} training (\num{26743} patches): used to train the \ac{CNN} to detect transition lines.
    \item \qty{10}{\percent} validation (\num{3343} patches): used after training to select the step corresponding to the best model (shown as a green star in Figure~\ref{fig:train-progress}).
    \item \qty{10}{\percent} testing (\num{3343} patches): used to evaluate the line classification accuracy after training.
\end{itemize}

Each patch is categorized as ``\emph{line}'' if at least one transition line annotation intersects the \numproduct{6 x 6} square at its center (pink squares in Figure~\ref{fig:patch-examples}a,b).
This labeling approach provides more context to the \ac{NN} while keeping the classification window narrow enough to fit between two transition lines.
Since the majority of the surface does not contain a line, only \qty{8.6}{\percent} (\num{3887}) of the patches are categorized as ``\emph{line}'', while the others are classified as ``\emph{no-line}''.

\section{Model Training}\label{sec:suppl-training}

\begin{figure}[H]
    \centering
    \includegraphics[width=.9\textwidth]{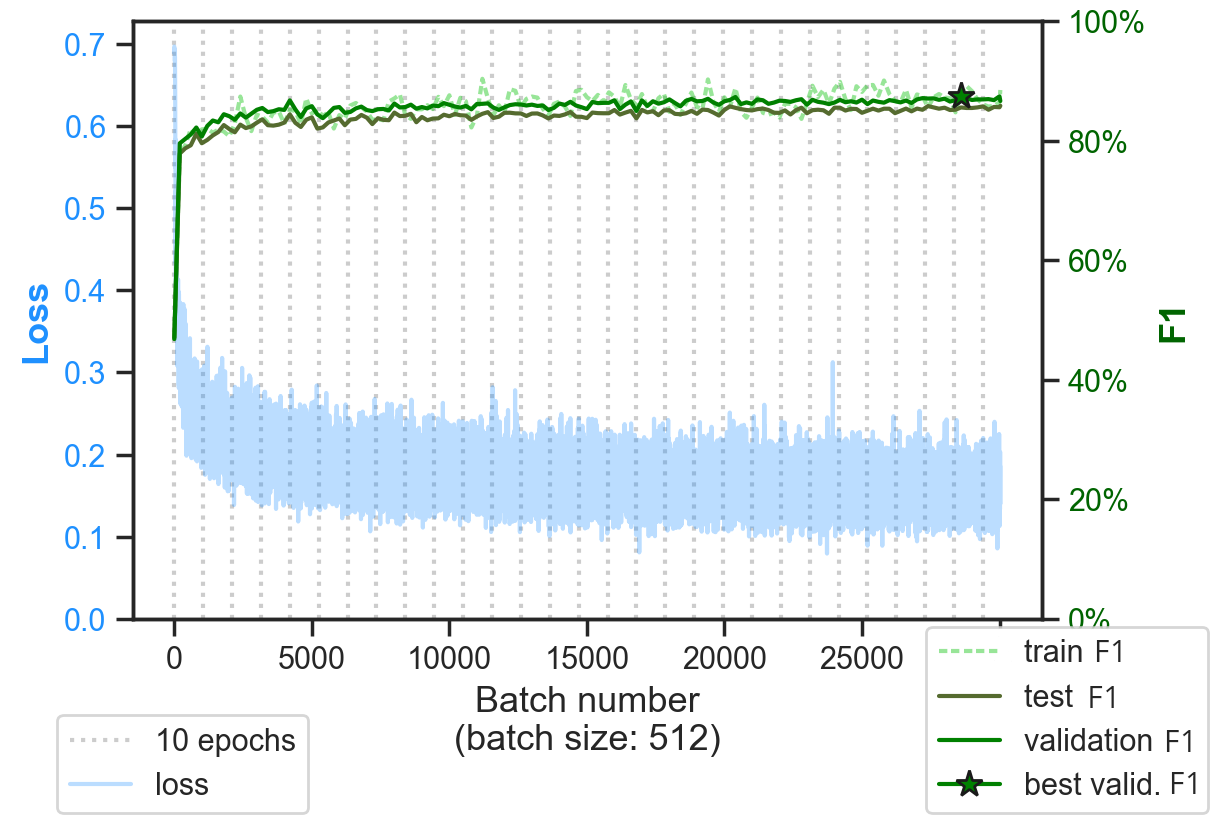}
    \caption[Example of \acl*{CNN} training progress]{
        Example of \acf*{CNN} training progress.
        The F1 score is the harmonic mean of precision and recall.
        The model performance quickly converges to a maximum, and no overfitting is visible.
    }
    \label{fig:train-progress}
\end{figure}

\begin{figure}[H]
    \centering
    \includegraphics[width=.4\textwidth]{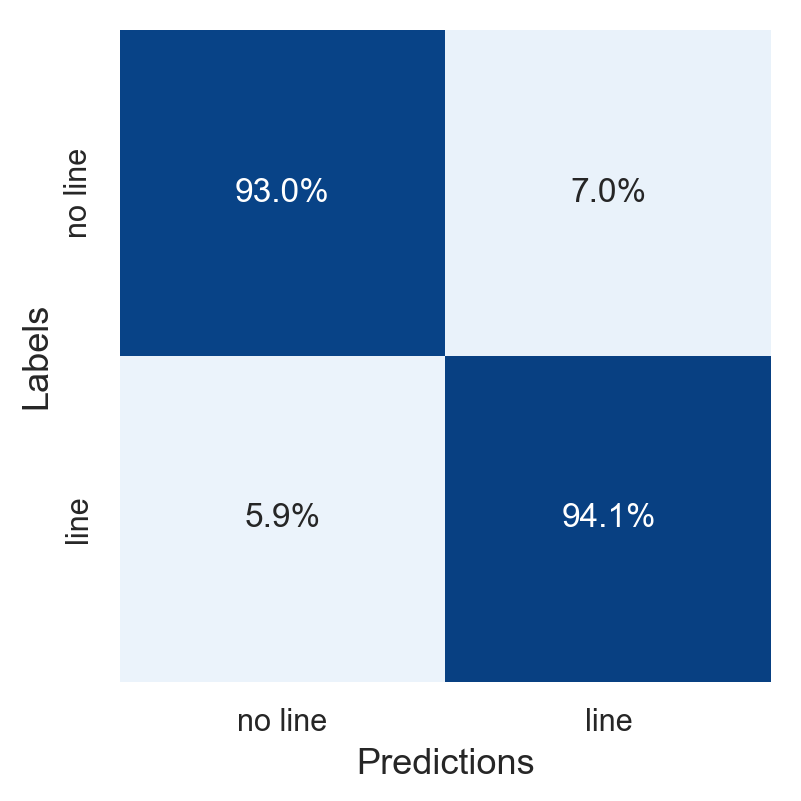}
    \caption[Example of a confusion matrix for a trained \acl*{CNN}]{
        Example of a confusion matrix for a trained \acf*{CNN}.
        On the test set, this model obtained the following scores: accuracy,~\qty{93.09}{\percent}; precision,~\qty{80.88}{\percent}; recall,~\qty{93.52}{\percent}; F1 score,~\qty{85.56}{\percent}.
    }
    \label{fig:confusion-matrix}
\end{figure}

The detection of transition lines is accomplished using \acp{CNN} containing \num{367267} free parameters (weights and bias), structured as presented in Table~\ref{tab:model-achitecture} and trained with the meta-parameters presented in Table~\ref{tab:meta-parameters}.
We did not observe overfitting during training, as shown in Figure~\ref{fig:train-progress}.
Once trained, the model reached satisfactory performance on the test set (example in Figure~\ref{fig:confusion-matrix}), where most of the errors were related to ambiguous labels or noise patterns.
The Python source code used for training and inference is available on \emph{GitHub}\footnote{Python source code: \href{https://github.com/3it-inpaqt/dot-calibration-v2}{github.com/3it-inpaqt/dot-calibration-v2}} and relies on \emph{PyTorch}~\cite{Paszke_2019}.

The meta-parameters listed in Tables~\ref{tab:model-achitecture} and~\ref{tab:meta-parameters} were determined through a grid search and informed guesses.
Some of them (input size, pixel size, and confidence threshold) are the result of a tradeoff between measurement time and classifier performance.
While further optimization and fine-tuning of these parameters could improve the classification accuracy, exploration efficiency, and overall calibration success, the primary aim of this study is to demonstrate the feasibility of the autotuning procedure in an experimental setting rather than achieving optimal performance.

The training of a \ac{NN} is partially stochastic due to the random initialization of the parameter values and mini-batch sampling during training.
To avoid selection bias, in this study, we trained a new \ac{CNN} using a different random seed for each autotuning run.
In a production context, it is possible to train and use only one \ac{CNN} to tune any device for a given \ac{QD} technology.

\begin{table}[H]
    \centering
    \caption[Architecture of the \aclp*{CNN} used to detect transition lines during the experiment]{
        Architecture of the \acfp*{CNN} used to detect transition lines during the experiment.
        A simplified schematic of the layer structure is represented in Figure~\ref{fig:setup}d.
    }
    \label{tab:model-achitecture}
    \begin{tabular}{r|l}
        \textbf{Layer}    & \textbf{Configuration}                       \\
        \hline
        Input             & \numproduct{18 x 18} pixels, \num{1} channel \\
        Convolution 1     & \numproduct{4 x 4} kernel, \num{6} channel   \\
        Convolution 2     & \numproduct{4 x 4} kernel, \num{12} channel  \\
        Fully connected 1 & \num{200} neurons                                  \\
        Fully connected 2 & \num{100} neurons                                  \\
        Output            & \makecell[l]{Binary classification           \\
                            (``\emph{line}'' or ``\emph{no-line}'')      \\
                            + uncertainty score}                         \\
    \end{tabular}
\end{table}

\begin{table}[H]
    \centering
    \caption[Meta-parameters of the \aclp*{CNN} used for training and inference.]{
        Meta-parameters of the \acfp*{CNN} used for training and inference.
    }
    \label{tab:meta-parameters}
    \begin{tabular}{r|l}
        \textbf{Meta-Parameter} & \textbf{Value}          \\
        \hline
        Number of train updates & \num{30000}             \\
        Loss function           & Binary cross-entropy    \\
        Optimizer               & Adam~\cite{Kingma_2014} \\
        Learning rate           & \num{0.001}             \\
        Dropout                 & \qty{60}{\percent}      \\
        Batch size              & \num{512}               \\
        Pixel size              & \qty{4}{\mV}            \\
        Confidence threshold    & \qty{90}{\percent}      \\
    \end{tabular}
\end{table}

\section{Confidence Score}\label{sec:suppl-confidence-score}

The confidence of model inference can be evaluated using several different methods~\cite{Gawlikowski_2021, Smith_2018, Guo_2017, Vaicenavicius_2019, Goan_2020, Kwon_2020}.
However, a previous study~\cite{Yon_2024} has shown that, in the context of \ac{QD} autotuning, a simple heuristic score based on the distance~\cite{Zaragoza_1998, Mandelbaum_2017} between the \ac{NN} output (denoted as $y$) and the inferred class value (see Formula~\ref{eq:conf-heuristic}) is an efficient approximation of the model's uncertainty.
This confidence score is leveraged to increase the tuning success rate with minimal overhead time by optimizing the exploration--exploitation tradeoff~\cite{Auer_2002, Simpkins_2008}.

\begin{equation}\label{eq:conf-heuristic}
    \text{Confidence score} = |0.5 - y| \times 2
\end{equation}

Based on offline tests, we empirically found that triggering the validation sequence if the confidence score was below \qty{90}{\percent} provided a good tradeoff between the success rate and exploration time.
Therefore, we used this value as a confidence threshold for the \num{20} online runs.
This value could be optimized during future experiments.

\section{Autonomous Exploration Strategy}\label{sec:suppl-exploration-strategy}

Each step of the exploration procedure (see Figure~\ref{fig:tuning-method}a)  consists in \emph{(i)} scanning a \numproduct{18 x 18} square region in the voltage space (referred to as ``patch''), \emph{(ii)} processing it through a \ac{CNN}-based line detection model, and \emph{(iii)} determining the next region to explore.

\begin{figure}
    \centering
    \includegraphics[width=.98\textwidth]{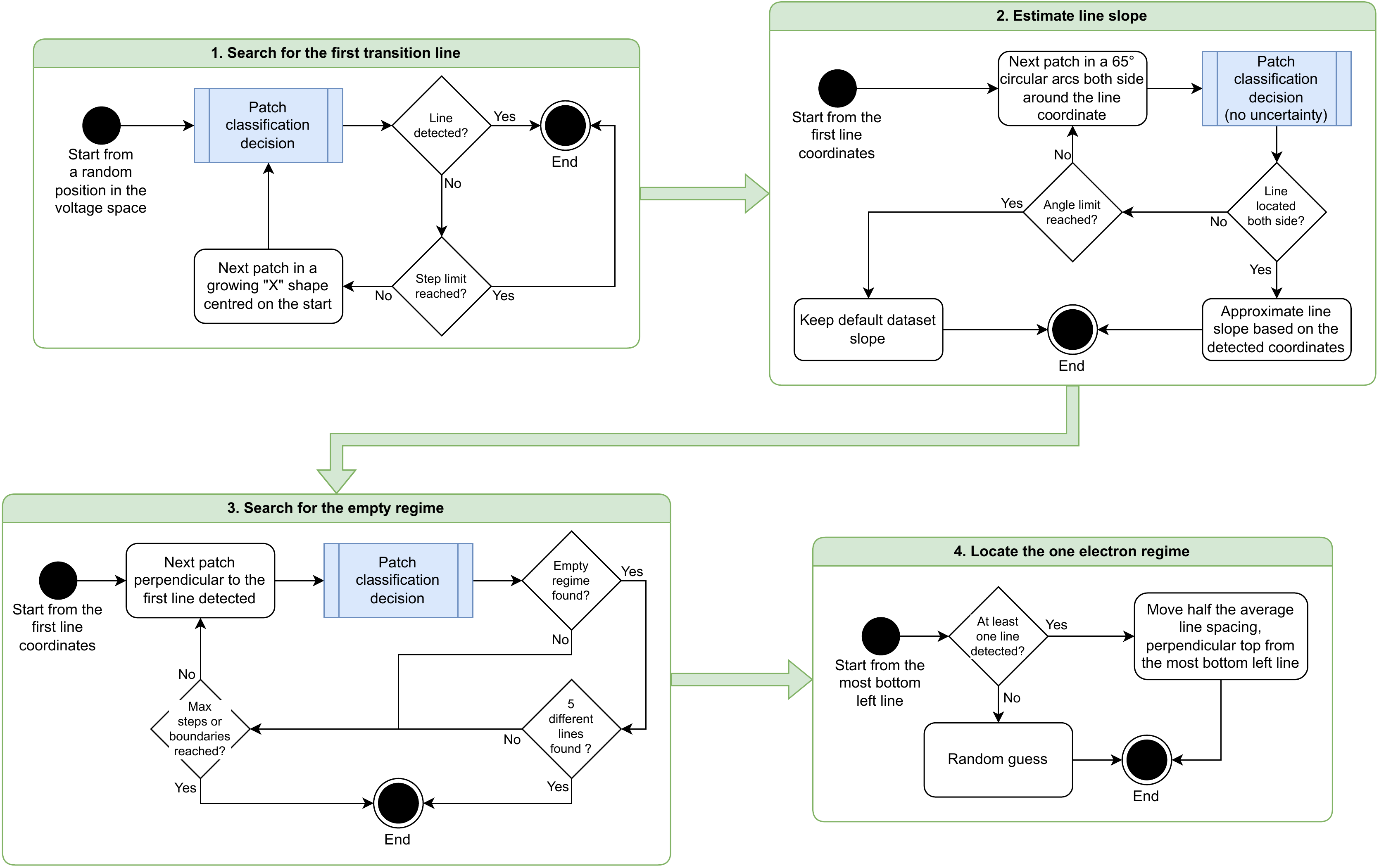}
    \caption{Flow diagram representing each stage of the autotuning exploration procedure.}
    \label{fig:exploration-flow}
\end{figure}

The coordinates for the next exploration step are determined by a four-stage exploration strategy (see Figure~\ref{fig:exploration-flow}):
\begin{enumerate}
    \item \textbf{Search for the first transition line:} The algorithm explores the voltage space in four directions, forming an expanding ``X'' pattern by following the diagonals from the starting point.
    This stage is successfully completed when the first transition line is detected with a confidence level above the threshold.
    \item \textbf{Estimate the line slope:} Patches are measured along two circular arcs centered on the first detected transition line.
    If the line is successfully located in both arcs, we use those two coordinates to estimate the slope of the line.
    If the line is not detected in one of the arcs, we keep the default slope value until the end of the procedure.
    \item \textbf{Search for the empty regime:} The exploration proceeds in the two directions perpendicular to the estimated line slope.
    This stage is successfully completed when the empty regime is detected (i.e., no transitionline detected for three times the average line distance in one direction) and at least five different transition lines are located.
    \item \textbf{Locate the one-electron regime:} Using the estimated line slope, the average distance between lines, and the position of the first line after the empty regime, we infer the gate voltages corresponding to the one-electron regime.
\end{enumerate}

The distance between two exploration steps is set to the width of the detection area (\qty{32}{\mV}), except in the second step, where eight steps are distributed in a \qty{65}{\degree}-arc on each side.
For further details on the exploration strategy and its benchmark on multiple offline datasets, see \citet{Yon_2024}.

\section{Run Statistics}\label{sec:suppl-runs-stats}

    \begin{table}[H]
        \centering
        \rowcolors{2}{}{lightgray!30}  
        \caption[Experimental run results]{Statistics and results of the \num{20} experimental runs.}
        \label{tab:runs-stats}
        \begin{tabular}{|r|c|c|c|c|c|c|}
            \hline 
            \textbf{Run ID} &
            \textbf{\#Steps} &
            \textbf{\makecell{Measurement\\Time (\unit{\second})}} &
            \textbf{\makecell{Processing\\Time (\unit{\second})}} &
            \textbf{\makecell{Total\\Time (\unit{\second})}} &
            \textbf{\makecell{Final Coord. \\(\unit{\volt}, \unit{\volt})}} &
            \textbf{\makecell{Charge\\Regime}} \\
            \hline 
            01     & \num{96}  & \integer{6363.42}    & \integer{259.124}  & \integer{6622.54} & (\round{-0.29789029535864975}, \round{0.8400000000000001}) & 1             \\
            02     & \num{71}  & \integer{4717.14}    & \integer{188.017}  & \integer{4905.15} & (\round{0.13502109704641352}, \round{0.756})               & 1             \\
            03     & \num{88}  & \integer{5860.31}    & \integer{233.776}  & \integer{6094.08} & (\round{0.1710970464135021}, \round{0.76})                 & 1             \\
            04     & \num{151} & \integer{10421.1}    & \integer{400.353}  & \integer{10821.5} & (\round{-0.27784810126582277}, \round{0.8280000000000001}) & 1             \\
            05     & \num{112} & \integer{7459.67}    & \integer{295.074}  & \integer{7754.75} & (\round{-0.0253164556962025}, \round{0.796})               & 1             \\
            06     & \num{167} & \integer{11462}      & \integer{445.129}  & \integer{11907.1} & (\round{-0.3018987341772152}, \round{0.8400000000000001})  & 1             \\
            07     & \num{72}  & \integer{4786.06}    & \integer{190.867}  & \integer{4976.93} & (\round{-0.29789029535864975}, \round{1.196})              & 4+            \\
            08     & \num{69}  & \integer{4670.28}    & \integer{182.135}  & \integer{4852.42} & (\round{0.23924050632911398}, \round{0.76})                & 1             \\
            09     & \num{165} & \integer{11474.8}    & \integer{445.305}  & \integer{11920.1} & (\round{-0.3018987341772152}, \round{0.8280000000000001})  & 1             \\
            10     & \num{117} & \integer{7792.2}     & \integer{309.014}  & \integer{8101.22} & (\round{0.17510548523206748}, \round{0.772})               & 1             \\
            11     & \num{76}  & \integer{5089.89}    & \integer{210.705}  & \integer{5300.6}  & (\round{0.2913502109704641}, \round{0.74})                 & 1             \\
            12     & \num{158} & \integer{10751.4}    & \integer{436.493}  & \integer{11187.9} & (\round{0.439662447257384}, \round{0.752})                 & 1             \\
            13     & \num{116} & \integer{7693.88}    & \integer{304.606}  & \integer{7998.48} & (\round{0.15105485232067517}, \round{0.776})               & 1             \\
            14     & \num{160} & \integer{11138}      & \integer{429.171}  & \integer{11567.2} & (\round{-0.3018987341772152}, \round{0.8360000000000001})  & 1             \\
            15     & \num{80}  & \integer{5299.03}    & \integer{211.177}  & \integer{5510.21} & (\round{-0.29789029535864975}, \round{0.8400000000000001}) & 1             \\
            16     & \num{89}  & \integer{5916.83}    & \integer{236.397}  & \integer{6153.23} & (\round{-0.08945147679324894}, \round{0.792})              & 1             \\
            17     & \num{116} & \integer{7790.36}    & \integer{310.934}  & \integer{8101.3}  & (\round{-0.0854430379746835}, \round{0.808})               & 1             \\
            18     & \num{62}  & \integer{4126.7}     & \integer{165.951}  & \integer{4292.66} & (\round{-0.0052742616033755185}, \round{0.784})            & 1             \\
            19     & \num{75}  & \integer{5063.24}    & \integer{199.675}  & \integer{5262.91} & (\round{-0.19367088607594934}, \round{0.812})              & 1             \\
            20     & \num{169} & \integer{11737.3}    & \integer{457.27}   & \integer{12194.6} & (\round{-0.3018987341772152}, \round{0.8280000000000001})  & 1             \\
            \hline 
        \end{tabular}
    \end{table}

\section{Patch Measurement Examples}\label{sec:suppl-patch-examples}

\begin{figure}[H]
    \centering

    \begin{subfigure}[t]{\textwidth}
        \centering
        \begin{subfigure}{0.18\textwidth}
            \centering
            \includegraphics[width=\textwidth]{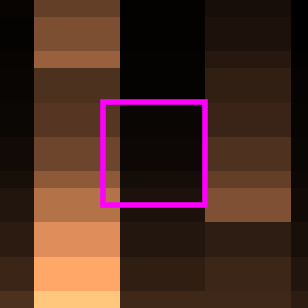}
        \end{subfigure}
        \begin{subfigure}{0.18\textwidth}
            \centering
            \includegraphics[width=\textwidth]{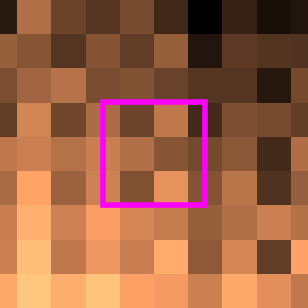}
        \end{subfigure}
        \begin{subfigure}{0.18\textwidth}
            \centering
            \includegraphics[width=\textwidth]{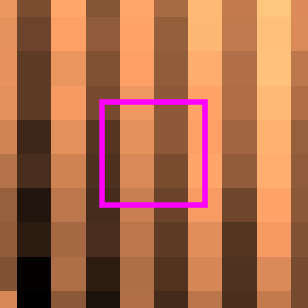}
        \end{subfigure}
        \begin{subfigure}{0.18\textwidth}
            \centering
            \includegraphics[width=\textwidth]{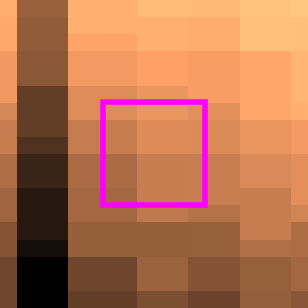}
        \end{subfigure}
        \caption{Four examples of offline patches labeled as ``\emph{no-line}''.}
    \end{subfigure}

    \begin{subfigure}[t]{\textwidth}
        \centering
        \begin{subfigure}{0.18\textwidth}
            \centering
            \includegraphics[width=\textwidth]{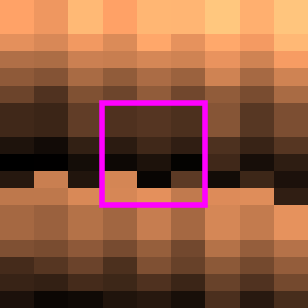}
        \end{subfigure}
        \begin{subfigure}{0.18\textwidth}
            \centering
            \includegraphics[width=\textwidth]{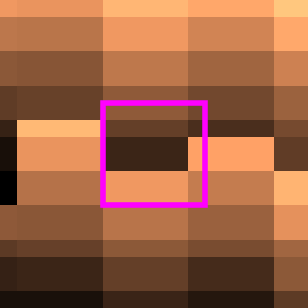}
        \end{subfigure}
        \begin{subfigure}{0.18\textwidth}
            \centering
            \includegraphics[width=\textwidth]{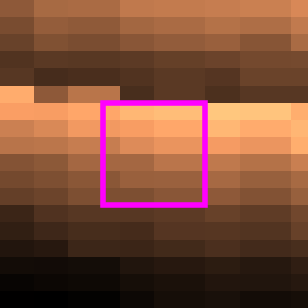}
        \end{subfigure}
        \begin{subfigure}{0.18\textwidth}
            \centering
            \includegraphics[width=\textwidth]{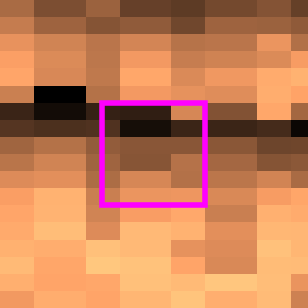}
        \end{subfigure}
        \caption{Four examples of offline patches labeled as ``\emph{line}''.}
    \end{subfigure}

    \begin{subfigure}[t]{\textwidth}
        \centering
        \begin{subfigure}{0.18\textwidth}
            \centering
            \includegraphics[width=\textwidth]{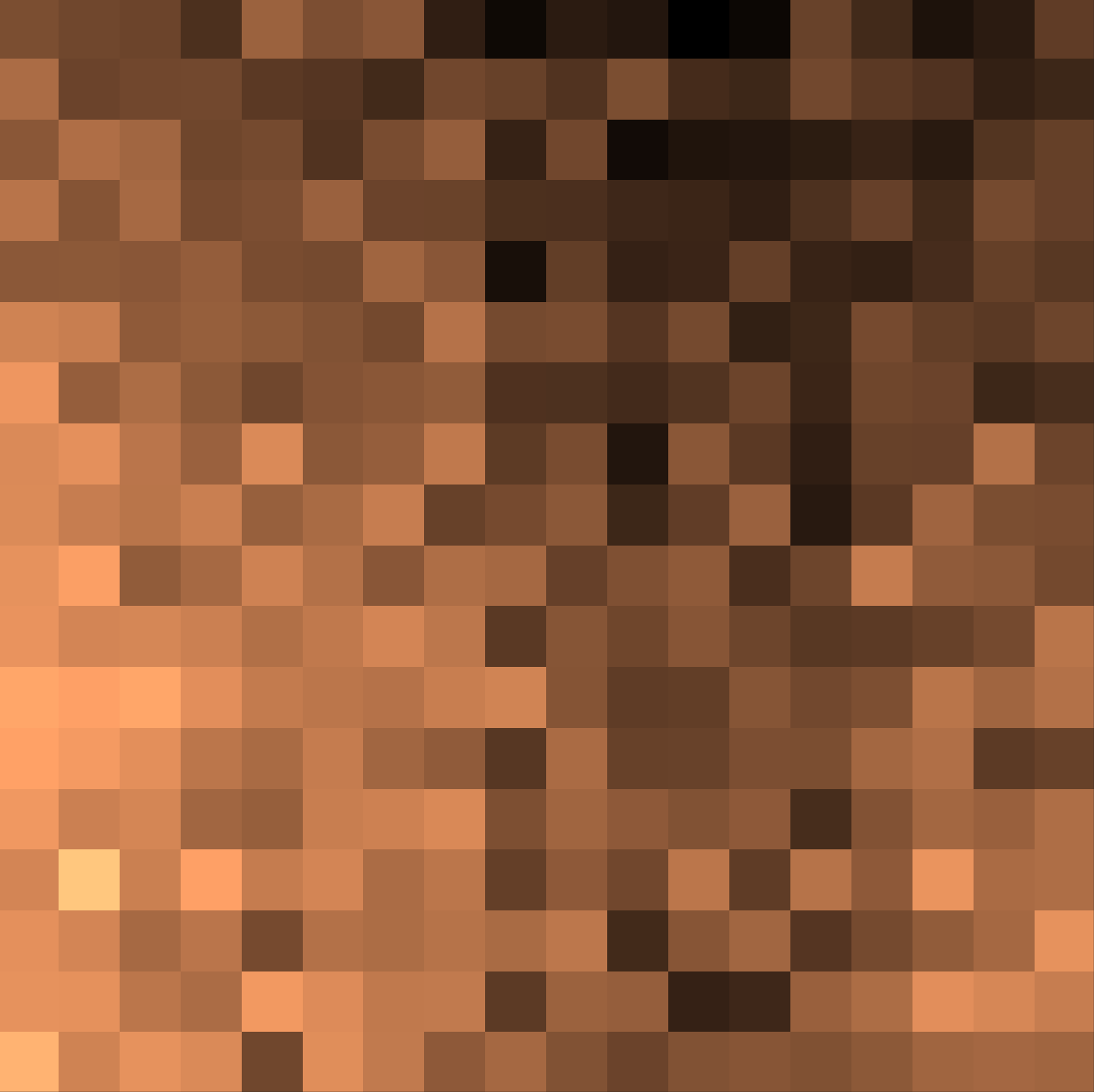}
        \end{subfigure}
        \begin{subfigure}{0.18\textwidth}
            \centering
            \includegraphics[width=\textwidth]{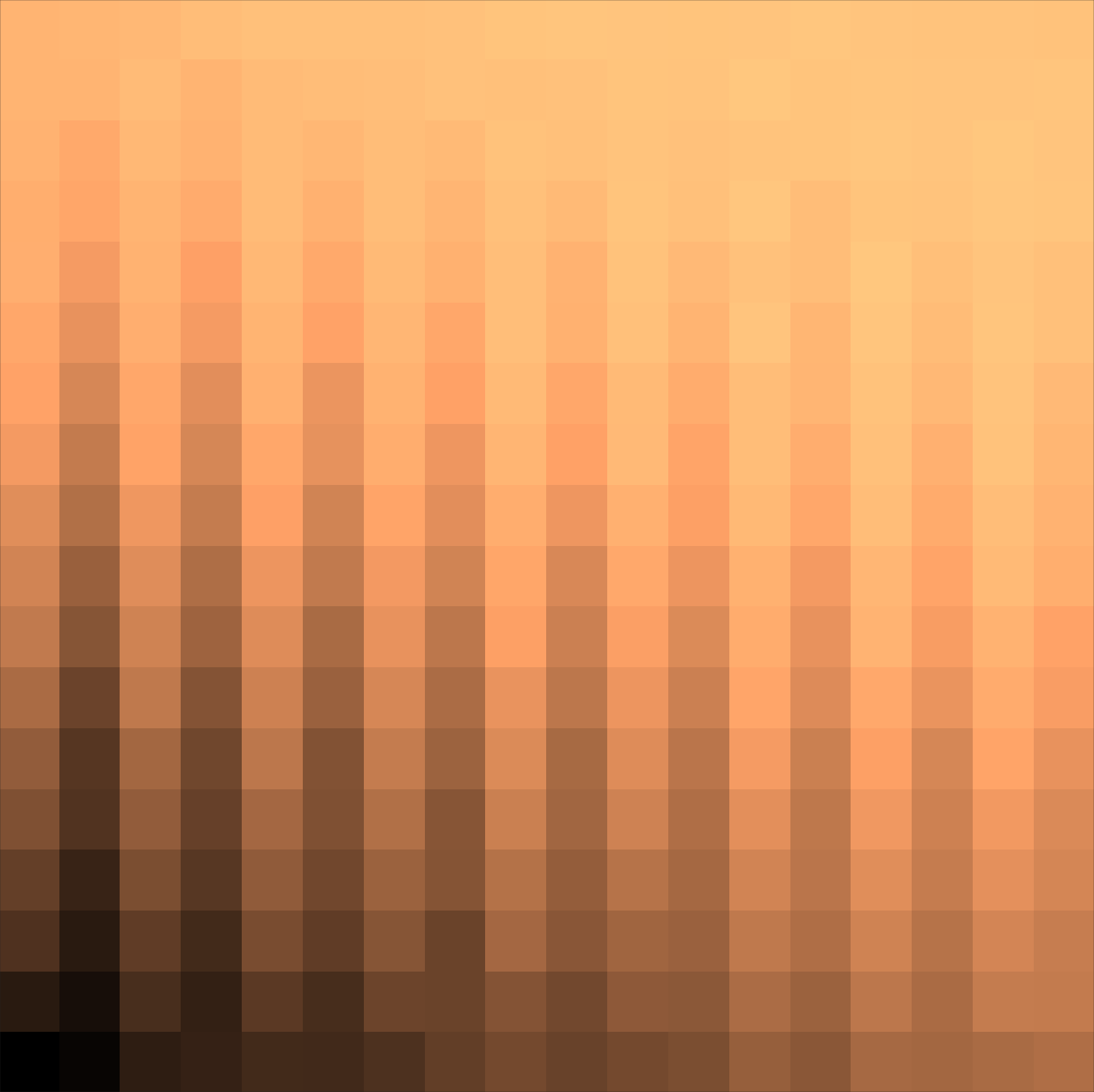}
        \end{subfigure}
        \begin{subfigure}{0.18\textwidth}
            \centering
            \includegraphics[width=\textwidth]{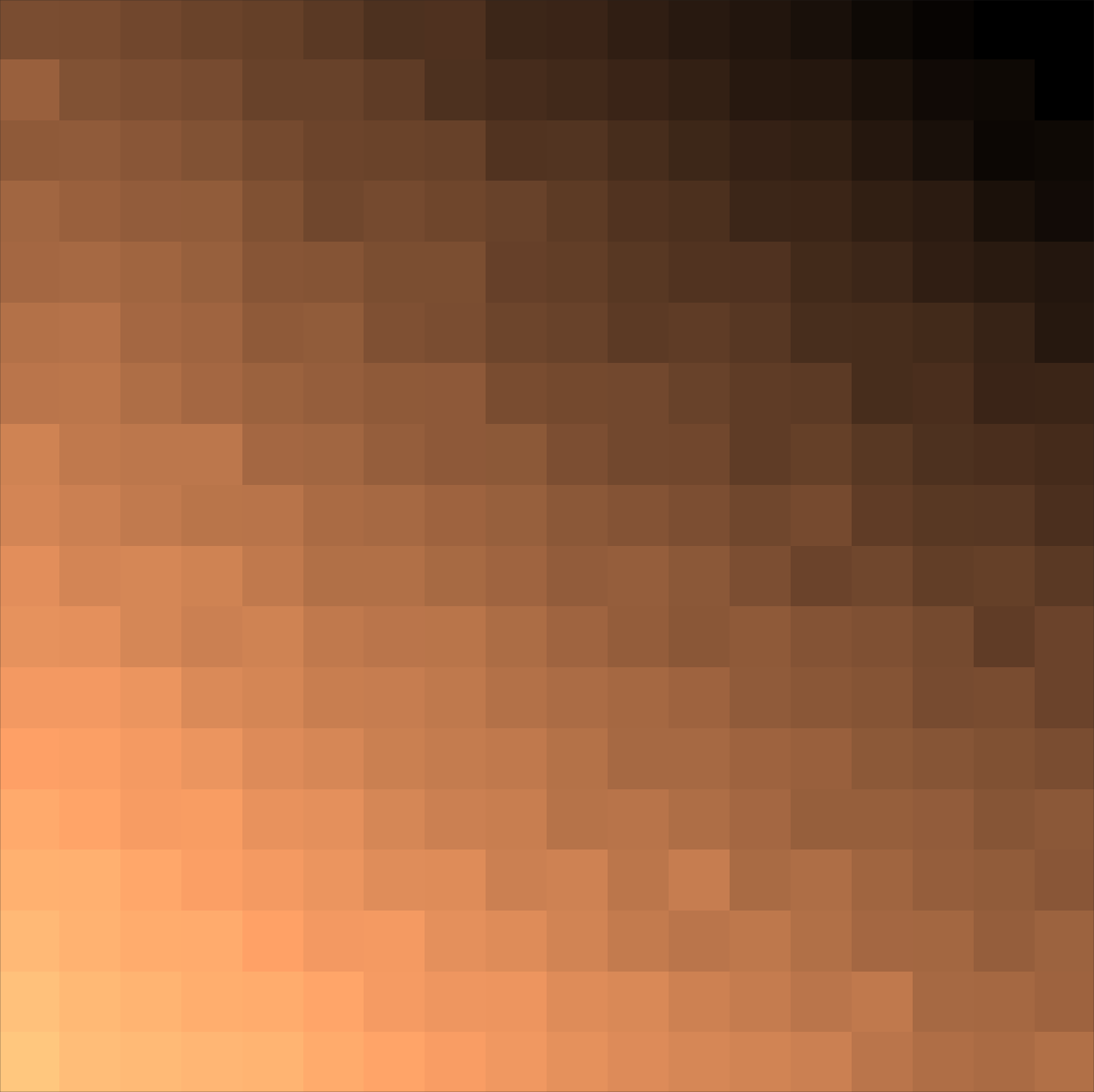}
        \end{subfigure}
        \begin{subfigure}{0.18\textwidth}
            \centering
            \includegraphics[width=\textwidth]{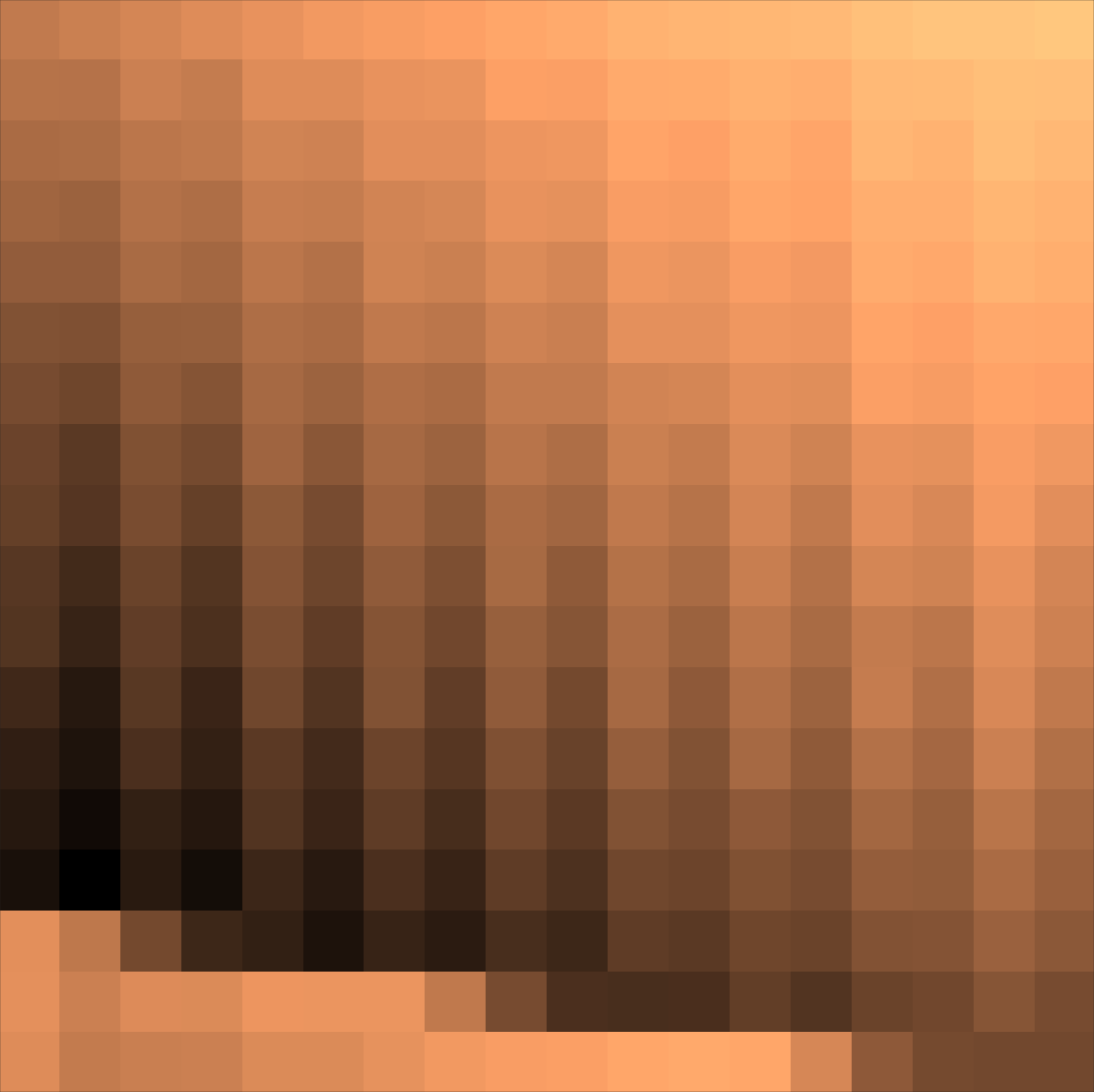}
        \end{subfigure}
        \caption{Four examples of online patches correctly classified as ``\emph{no-line}''.}
    \end{subfigure}

    \begin{subfigure}[t]{\textwidth}
        \centering
        \begin{subfigure}{0.18\textwidth}
            \centering
            \includegraphics[width=\textwidth]{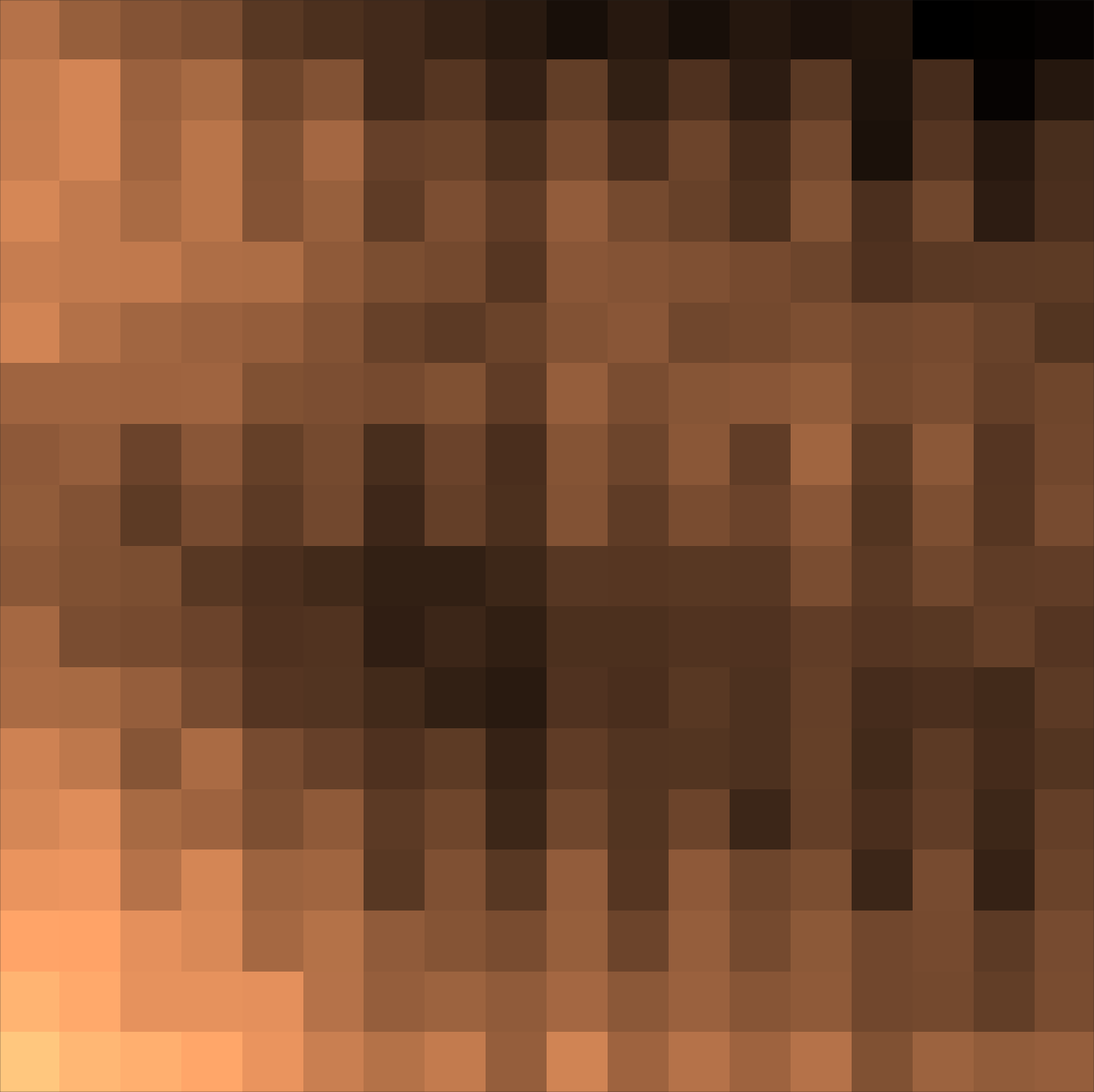}
        \end{subfigure}
        \begin{subfigure}{0.18\textwidth}
            \centering
            \includegraphics[width=\textwidth]{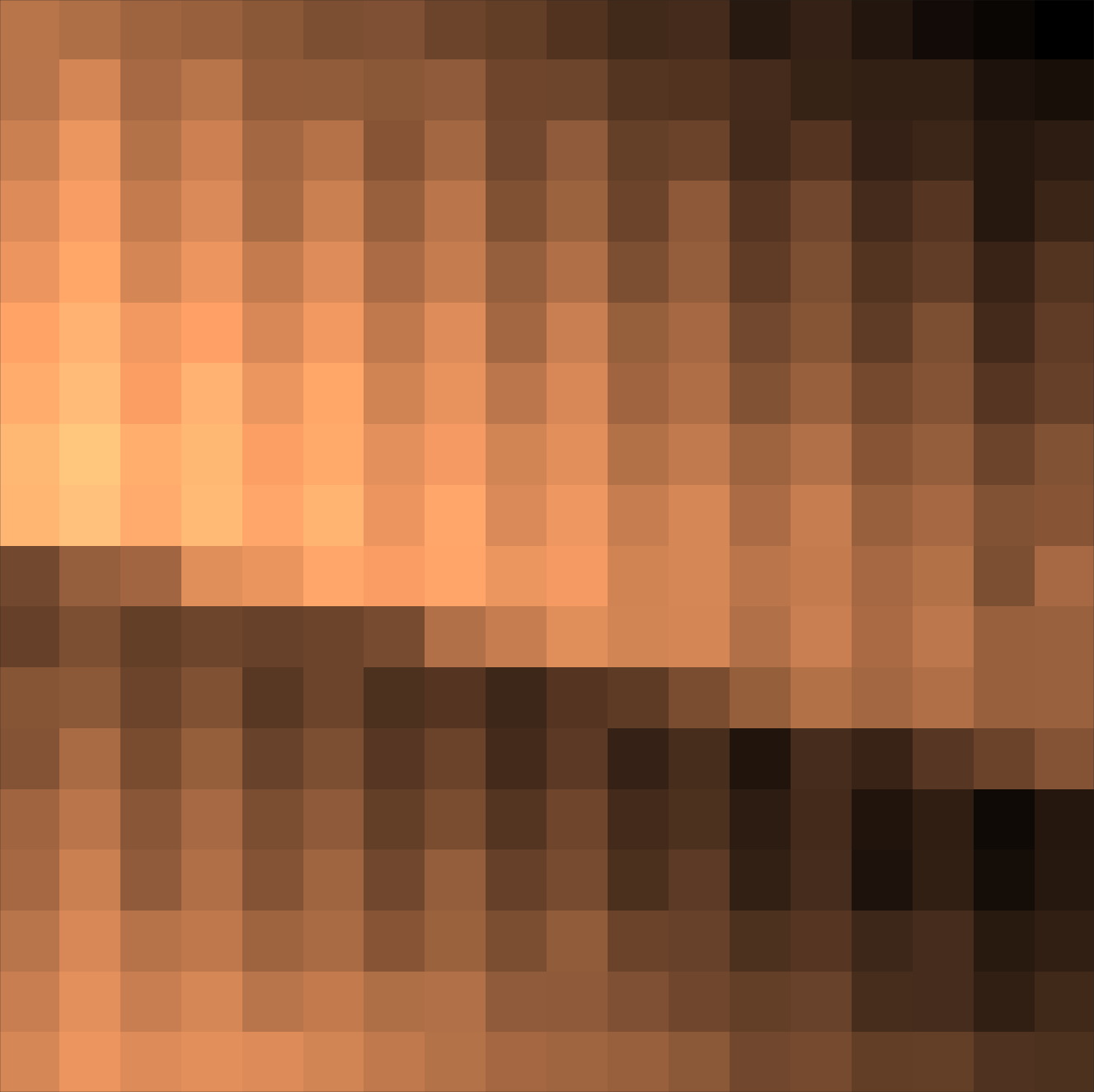}
        \end{subfigure}
        \begin{subfigure}{0.18\textwidth}
            \centering
            \includegraphics[width=\textwidth]{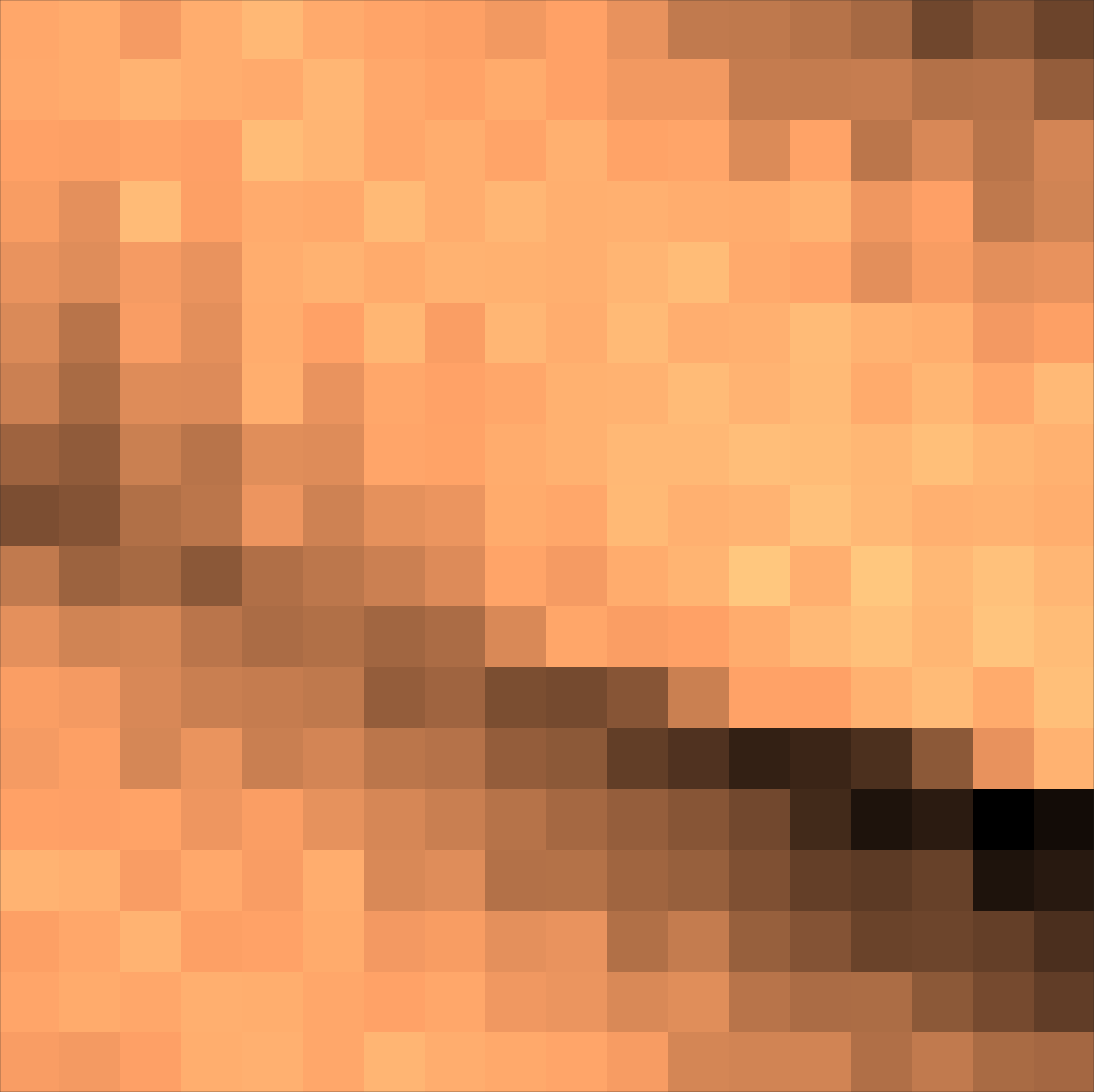}
        \end{subfigure}
        \begin{subfigure}{0.18\textwidth}
            \centering
            \includegraphics[width=\textwidth]{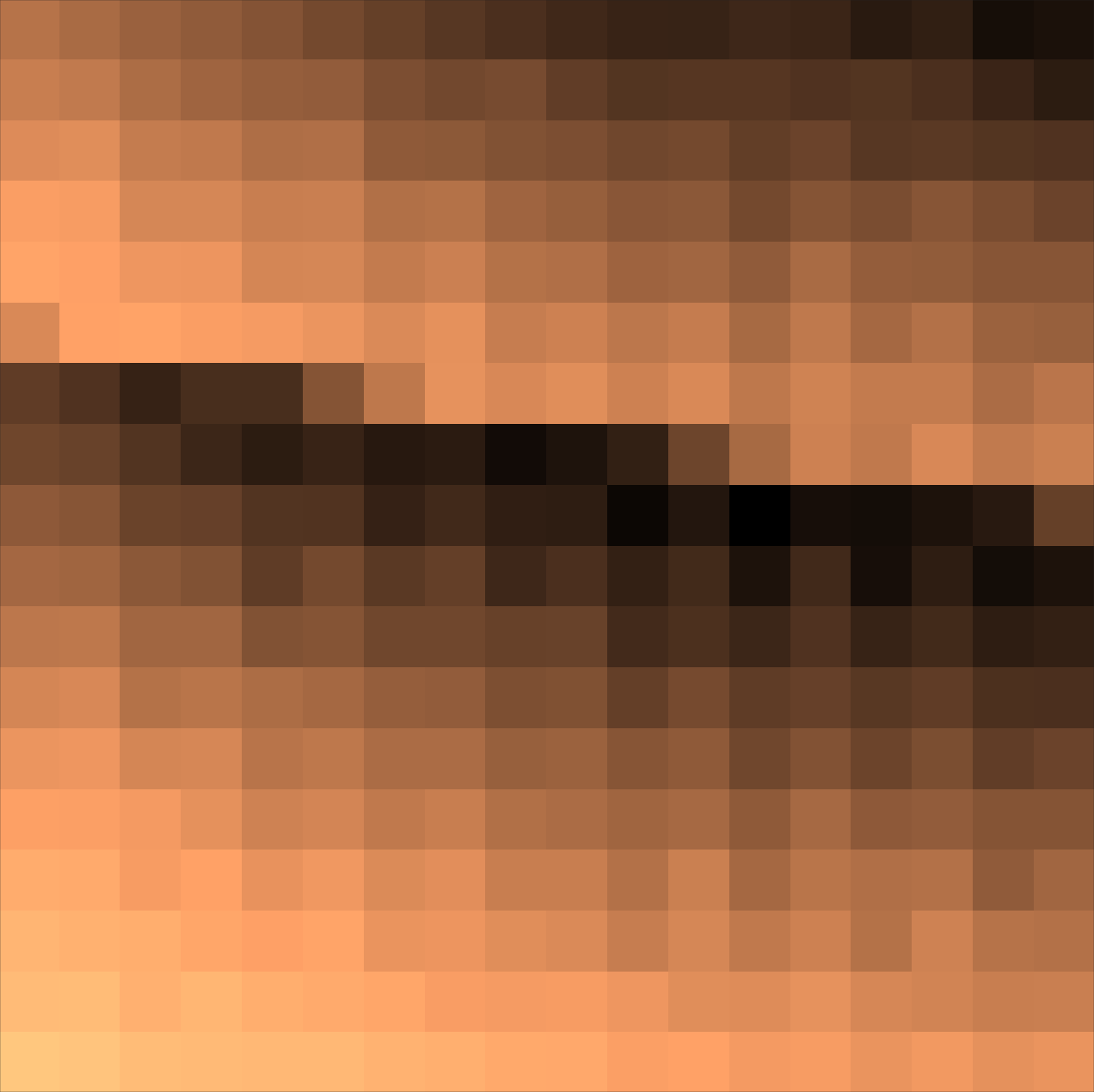}
        \end{subfigure}
        \caption{Four examples of online patches correctly classified as ``\emph{line}''.}
    \end{subfigure}

    \caption[Patch samples for each class from the offline training set and the online experiment]{
        Patch samples for each class from the offline training set and the online experiment.
        \textbf{a,b)}~The patches are measured using variable step sizes depending on the stability diagram, then interpolated at \qty{4}{\mV} per pixel.
        A patch is labeled as ``\emph{line}'' only if a transition line annotation intersects with the detection area (highlighted in pink) at its center.
        \textbf{c,d)}~The online patches are measured using a step size of \qty{4}{\mV} per pixel.
    }
    \label{fig:patch-examples}
\end{figure}

It was necessary to interpolate the images of the offline stability diagrams before using them to train the \acp{CNN} due to the variable sweeping step size used during the original experiments.
This preprocessing slightly affected the quality of the offline training by duplicating some pixels (see patch examples in Figure~\ref{fig:patch-examples}a,b).
On the other hand, the patches measured during the online experiment did not suffer from this limitation, as it was possible to define the desired step size in the measurement interface.
This led to a better resolution of the patch, with clearer lines (see patch examples in Figure~\ref{fig:patch-examples}c,d).

\end{document}